\documentstyle[aps,pre,psfig]{revtex}
\newcommand{\beq}{\begin{equation}}
\newcommand{\eeq}{\end{equation}}
\newcommand{\beqa}{\begin{eqnarray}}
\newcommand{\eeqa}{\end{eqnarray}}
\begin{document}
\twocolumn[\hsize\textwidth\columnwidth\hsize\csname@twocolumnfalse\endcsname
\title{Long-term properties of time series generated by
a perceptron with various transfer functions}

\author{Avner Priel and Ido Kanter}
\address{Minerva Center and Department of Physics, 
	Bar-Ilan University, 52900 Ramat-Gan,
Israel}

%\date{\today}

\maketitle

\begin{abstract}
We study the effect of various transfer functions on the properties 
of a time series generated by a continuous-valued feed-forward network 
in which the next input vector is determined from past output values. 
The parameter space for monotonic and non-monotonic transfer functions
is analyzed in the unstable regions with the following main finding;
non-monotonic functions can produce robust chaos whereas monotonic functions
generate fragile chaos only. In the case of non-monotonic functions, 
the number of positive Lyapunov exponents increases as a function of one 
of the free parameters in the model, hence, high dimensional chaotic 
attractors can be generated. We extend the analysis to a combination 
of monotonic and non-monotonic functions.
\end{abstract}

\vspace{.6cm}
]

\section{Introduction}
One of the developing  subjects in the research of neural networks
is the analysis of time series. There are several approaches in this field
such as prediction, characterization, modeling etc. 
In this paper, we focus on understanding the interplay between the type
of transfer function used and some quantitative measures of the time series
generated. In particular we are interested
in the classification of the possible types of sequences generated by the 
network and their characteristics according to the nature of the attractor 
of the dynamics. 
Previous analytical studies concentrated on the stable regime of the parameter
space of feed-forward networks with a feedback loop that generate 
time series \cite{bg,prl,jpa,nips,martin,sgen_mln}. 
One of the main questions we address
in this paper regards the behavior of the system in the {\it unstable} 
regime and how varying the transfer function affects the
asymptotic behavior of the sequence generated by the model.

In order to characterize the dynamical system we analyze its
properties while varying some control parameters.
Analysis of the parameter space of a map
enables us to classify the type of flow in phase space in the vicinity of 
a given vector of parameters.
An interesting question which arises is whether it is possible to
generate high dimensional attractors and control their properties, e.g.\ ,
the attractor dimension (a global property of the phase space) and the 
robustness (a local property of the parameter space).

As we shall see, there exists a clear distinction between monotonic
and non-monotonic transfer functions, e.g.\, in terms of the structure of 
parameter space and attractor dimension. We shall try to illuminate this
phenomenon as well as its relation to the possibility of generating 
robust chaos. The concept of robust chaos (see \cite{soumitro}) 
is associated with an attractor
for which the number of positive Lyapunov exponents (in a region of parameter
space) is larger than the number of free (accessible) parameters in the model.
Moreover, in the vicinity of a chaotic parameter's vector, no periodic 
attractors are found. We give strong indications to support our 
conjecture which states that monotonic functions are not capable of 
generating robust chaos while non-monotonic functions are.

The analysis presented in this paper is performed mostly
for a perceptron with weights composed of 
a single Fourier component with an additional bias term, except where 
otherwise mentioned. There are two main reasons behind this choice. 
First, this choice leads to a two dimensional parameter space ($\beta,b$),
gain and bias respectively, which can be conveniently visualized, 
whereas larger parameter space is more difficult to handle. Second, all the 
important characteristics are already manifested in this case.
The existence of a bias
in the weights is important for producing unstable dynamics in the case of
monotonic functions, therefore it is crucial to include this parameter.

The paper is organized as follows. In section \ref{sec_model}, the
model is described and some of the relevant results previously obtained are
reviewed. In section \ref{sec_mon_fun} the class of monotonic functions, 
such as the hyperbolic-tangent function, is analyzed. 
The cases of two and three input units,
$N=2,3~$, are examined and compared to previous findings. We analyze the 
parameter space using numerical methods, e.g.\, calculating Lyapunov spectrum, 
attractor dimension (see Appendix), to identify stable and chaotic regions. 
The main conclusion is that the model with monotonic functions is indeed 
rather stable in the sense that even in regimes where chaotic behavior can
be found, the chaos is fragile and small variations of the parameters drive
the system to stable dynamics. 
In section \ref{sec_nonmon_fun} non-monotonic functions are examined both
analytically and numerically.
The issue of high dimensional attractors is treated as well as the structure
of parameter space and the possibility of generating robust chaos.
Finally, in section \ref{sec_comb} we discuss the case of a transfer 
function which is a combination of monotonic and non-monotonic functions.
The appendix contains the technical details concerning our analysis of the
parameter space which one should refer to while reading sections
\ref{sec_mon_fun} and \ref{sec_nonmon_fun}.

\section{The Model}\label{sec_model}

Let us consider a perceptron with $N$ input units and weights 
$\vec{W}$. For a given input vector at time step $t$,
${\vec S}^t$ ($S_j^t , j=1,\ldots,N$), the network's output $S_{out}^t$ 
is given by
\beq\label{s_out}
S_{out}^t=f \left( \beta \sum_{j=1}^N  W_j  S_j^t \right)
\eeq
\noindent
where $\beta$ is a gain parameter and $f$ is a transfer function.
The input vector at time $t+1$ is defined by the previous output values in the 
following dynamic rule:
% (figure \ref{fig_model})
\beq\label{dyn_rule}
S_1^{t+1} = S_{out}^t \quad ; \qquad S_j^{t+1}=S_{j-1}^t
\quad j=2, \ldots ,N
\eeq

\noindent
Since the network generates an infinite sequence from an initial state, 
this model was denoted in previous papers as a Sequence-Generator (SGen)
e.g.\ \cite{prl}.
We restrict the discussion to bounded, symmetric nonlinear transfer functions, 
i.e.\, 
\beq
f:{\cal R} 
\rightarrow {\cal R}  ~, ~ |f(x)|< \infty \quad \forall x \in {\cal R} .
\eeq

\noindent
A prescription for the weights is given by the following form

\begin{eqnarray}\label{weights_general}
W_j  =  \sum_p a_p \cos({2\pi \over N} k_p j + \pi \phi_p) + b \nonumber \\ 
j=1 \ldots N , \quad \phi_p \in [-1..1] 
\end{eqnarray}

\noindent
where $\{ a_p \}$ are constant amplitudes; $\{ k_p \}$ are positive integers 
denoting the wave numbers; $b$ is the bias term and $p$ runs over the number 
of Fourier components composing the weights. 
In the following we investigate only the cases $p=1$ or $2$ in order to 
keep the dimension of the parameter space as small as possible.

Our main concern is the differences imposed by the transfer function on the
asymptotic behavior of the time series generated by the model. We 
concentrate on two classes of functions, monotonic and non-monotonic,
which are exemplified in detail by hyperbolic-tangent and $\sin$ functions
respectively. 
This model was analyzed in various cases, all of them in the stable regime,
for moderate values of the gain parameter $\beta$. The other extreme, 
$\beta \rightarrow \infty$ was also treated (see \cite{bg,martin}). We 
concentrate on the intermediate regime for which unstable behavior emerges.

Let us review the relevant results previously obtained.
In the case of a `perceptron-SGen' with general weights and an odd 
transfer function, the system undergoes a Hopf 
bifurcation at some critical value of the gain parameter. The stationary 
solution above the bifurcation value is characterized by a quasi-periodic 
attractor flow governed by 
one of the Fourier components of the power spectrum of the weights,
hence the attractor dimension ($AD$) is typically one.
This type of flow becomes unstable at higher gain value, this being
the focus of this paper.
We should point out here that there are cases for which 
a stable two-dimensional $(2D)$ attractor is observed \cite{martin}, 
nevertheless their measure is zero.  
The results were extended to Multi-Layer Networks
in \cite{prl,sgen_mln} where the attractor dimension, in the stable regime, 
is found to be bounded in the generic case by the number of hidden units,
independent of the complexity of the weight vectors.

\section{Monotonic functions}\label{sec_mon_fun}

In this section we discuss the case of monotonic transfer functions. This
family of functions is typical to neural networks for several reasons, one 
of which is their biological plausibility (see e.g.\ \cite{hertz}). 
For the rest of this section we use the hyperbolic-tangent function 
as being representative; however, the main results are common to other
monotonic transfer functions.
The output, $S_{out}$, for this case is given by
\beq\label{tanh_output}
S_{out}^t=\tanh(\beta \sum_{j=1}^N W_j S_j^t )
\eeq

\noindent
The weights consist of a single biased Fourier component as follows

\beq\label{weights_1p}
W_j = a \cos({2\pi \over N} k j + \pi \phi) + b \qquad j=1 \ldots N , \quad
\phi \in [-1,1]
\eeq

\noindent
In the following we set $a=1$ to reduce the dimensionality of the parameter
space.
For the simplest case of two inputs, $N=2$, 
the equation which describes this map is simply:
\[ S^{t+1}= \tanh \left[ \beta (W_1 S^t + W_2 S^{t-1}) \right] \]
\noindent
where $W_i ~ i=1,2$ are given by Eq.\ (\ref{weights_1p}). 
The special case $~\phi=1~$ reduces to
\beq\label{eq_sgen_N2}
S^{t+1}= \tanh \left[ \beta \left( S^t (1+b) + S^{t-1} (-1+b) \right) \right] 
\eeq

\noindent
which is equivalent to a physical model of a magnetic
system, ANNNI model \cite{annni,annni_explain}, that was intensively 
investigated in the past. 
This map is capable of generating stable attractors (nontrivial fixed 
points, periodic and quasi-periodic orbits) as well as unstable chaotic 
behavior. The commensurate phase of the map is presented in \cite{annni}
and therefore will be omitted here. 

\begin{figure}
\centerline{\psfig{figure=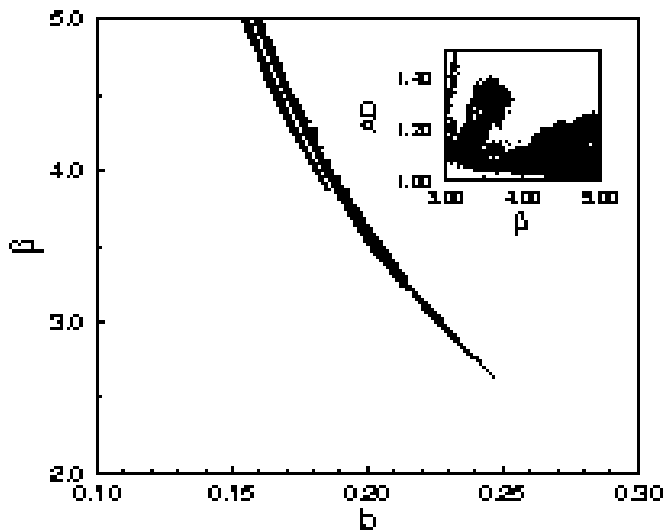,height=10.2cm}}
\caption{Analysis of a region in parameter space for hyperbolic-tangent
transfer function and $N=2$. Points that lead 
to chaotic trajectories are marked. 
The remaining space in this region leads to stable attractors.
The center of the circle at $(4.24,0.178)$ represents the chaotic point 
discussed in \protect\cite{annni} . 
Insert: the attractor dimension (AD) of the chaotic 
points shown in this region. For each value of the parameter $\beta$, the 
AD of all the chaotic points along the $b$ axis are drawn.
}
\label{fig_th_N2}
\end{figure}

The goal of the analysis of the parameter space is to classify the dynamics 
in a region of parameter space. 
The analytical part is unfortunately absent here due to the limitations
set by the transfer function.
It transpires that the class of monotonic functions does not give rise to
critical points since the first order derivatives of the map are always
positive, therefore the determinant is bounded away from zero. For large
periodic orbits, however, it can come very close to zero, therefore 
the structure can be somewhat more similar to maps that
do contain critical points (see discussion in section \ref{sec_nonmon_fun}).
Therefore, let us turn to a numerical analysis.
In order to answer questions such as the existence of a robust chaos, 
the parameter space is sampled in a high resolution, up to $10^{-5}$ in each 
direction. Figure \ref{fig_th_N2} depicts a section of the parameter space
for the case $\phi =1$. The black area leads to chaotic behavior 
with one positive Lyapunov exponent. This area is not a compactly dense 
unstable region. In fact each unstable point has stable neighbors which lead
to periodic attractors.
The remaining space in this region gives rise to stable attractors. 
The center of the bold circle represents the point mentioned in \cite{annni} 
which leads
to chaotic behavior with the parameter value $[\beta=4.24155, b=0.17881]$. 
A perusal of the figure shows that the unstable region is constructed
around a $1D$ curve. For other choices of the phase $\phi$ the 
parameters of this curve change, not its nature.
Moreover, we note that the unstable points lead to a 
mixed behavior, i.e.\, both stable and unstable behavior can be obtained 
from the same vector of parameters, depending on initial conditions. 
The dimension of the chaotic attractors in this
region was calculated using the Kaplan-Yorke conjecture \cite{kap_yorke} 
(see Appendix) and presented in the insert of Fig.\ \ref{fig_th_N2}. 
The figure is a projection of the $3$ variables, $AD, \beta, b$, 
on the $AD$-$\beta$ plane, i.e.\, for each value
of the gain $\beta$ all the unstable points along the $b$ axis are presented.
It is clear that the dimension is typically between $1.0$ and $1.3$ 

% Figure 1 here

\begin{figure}
\centerline{\psfig{figure=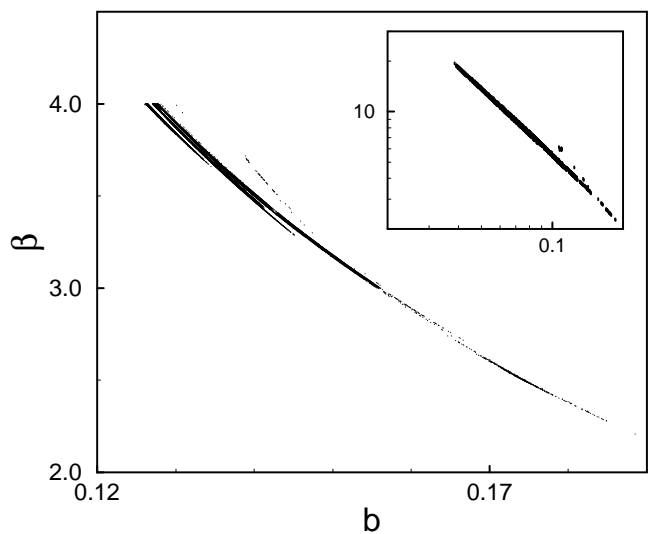,height=7.2cm  }}
\caption{A region in parameter space for hyperbolic-tangent transfer function 
and $N=3$. Points that lead to chaotic trajectories are marked. 
The remaining space in this region leads to stable attractors. Insert: 
A continuation of the main figure for smaller values of $b$ 
in a $\log$-$\log$ plot.
}
\label{fig_th_N3}
\end{figure}

The same analysis was applied to the case $N=3$ (which is similar to the 
dynamics of the ANNNI model with competing interactions 
between third neighbors along the axial direction \cite{moreira}). 
Figure \ref{fig_th_N3} presents the results of the same analysis which was
applied for $N=2$. The insert shows the continuation of the figure for higher
gain values indicating that the unstable behavior can be found for any small 
bias values. 
The reason for taking $\phi=1$ here as well, originates from the fact that
larger phases tend to generate more unstable regions.
A similar behavior was found in the case of a binary output 
($\beta \rightarrow \infty$) where the size of the cycles increases 
with $\phi$ \cite{bg}.
It was found that the phase diagram is basically the same as for $N=2$,
i.e.\, the main features as 
described above are also present here. 
%
% Figure 2 here
%

The general case $\phi<1$ can be analyzed in the same manner as that 
described above. We note that for $N=2$ no unstable regions are found 
for $\phi<{1 \over 2}$. \\

\begin{figure}
\centerline{\psfig{figure=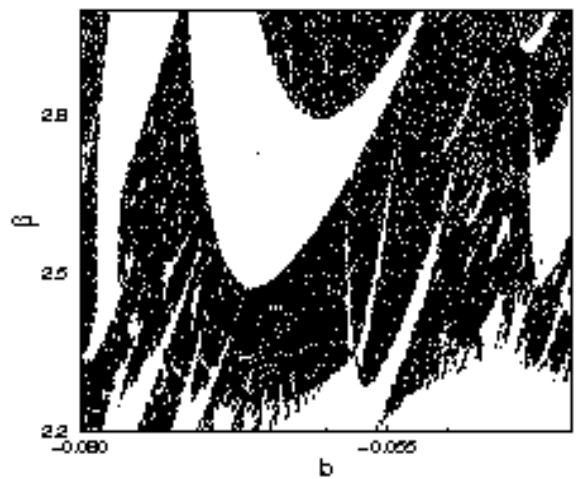,height=7.2cm  }}
\caption{Example of a region in parameter space for hyperbolic-tangent 
transfer function and $N=9$ where points that lead to chaotic trajectories
are marked. The remaining space in this region leads to stable attractors.
}
\label{fig_th_N9}
\end{figure}

In higher dimensions (larger $N$) it is necessary to use more Fourier 
components to describe general weights, therefore more parameters
are required - amplitudes and phases. As before, we restrict the dimension
of parameter space to two.
Figure \ref{fig_th_N9} depicts a region in parameter space for the case 
$N=9$ with $\phi=1$ (unstable behavior is found outside this region as well).
The unstable points cover a significant part of the space.  
Qualitatively, the parameter space is similar 
to that of $N=2,3$ in the sense that the chaotic regions are mixed
and fragile. However, as $N$ increases the structure of the parameter space
becomes more involved as larger cycles become available.
Moreira and Salinas \cite{moreira} have already mentioned that such a 
complication is expected at larger $\beta$ in their model ($N=3$). 
We should stress here that the apparent dense regions of unstable points 
{\bf do not} imply robust chaos since all the characteristics discussed
previously for smaller systems are present here, namely there is only a single
positive Lyapunov exponent; the chaotic regions are fragile, i.e.\, in the 
vicinity of every unstable point there exists a stable one.
In particular, these points generate a mixed behavior in phase space. 
Stable and unstable attractors are possible, depending on the initial 
conditions, hence both have a non-vanishing basin of attractions.

% Figure 3 here

The examples provided so far consist of weights with a single Fourier
component.
Nevertheless, we do not expect any significant quantitative changes in the
cases where the weights consist of more Fourier components,
besides the obvious addition of free parameters. 
The reason is that the number of positive
Lyapunov exponents does not increase.
For conciseness, we tested the case of two Fourier components with bias,
$p=2$ in Eq.\ (\ref{weights_general}). A few cases with arbitrary amplitudes
and phases were chosen. The results indicate 
that our conclusions are applicable in the more general case. 

In all our simulations we found no regions
with more than a single positive exponent (for $N$ up to $60$), including
many cases with randomly chosen weights.
Therefore, we conjecture that the SGen with hyperbolic-tangent function 
typically exhibits unstable behavior with a single positive Lyapunov exponent. 

We conclude with the observation that the bias term $b$, Eq.\ 
(\ref{weights_general}) is
crucial for producing chaotic behavior in the model with a monotonic 
transfer function. Another important ingredient is the existence of a large
enough phase, at least when the weights consist of a single Fourier 
component. It is possible that
additional Fourier components are sufficient to generate unstable 
behavior (without large phase), however, larger phase 
significantly increases the number of unstable points.

\section{Non-monotonic functions}\label{sec_nonmon_fun}

Applying a non-monotonic transfer function dramatically alters the structure
of parameter space with respect to monotonic functions. 
One is able to observe {\it robust chaos}, and the possible number
of positive exponents is no longer bounded by one. 
In the following analysis we treat the class of odd non-monotonic functions 
and use the '$\sin$' as a representative function. 

As mentioned in section \ref{sec_model}, quasi-periodic stationary 
solutions were found analytically for odd functions 
which are valid below some critical value of the gain parameter. 
In this section we focus on the region beyond that value. Note 
that in contrast to monotonic functions, unstable dynamics can be obtained
with phase and bias equal to zero. Indeed, in the sequel we use 
$\phi=0$ and only two dimensional parameter space,
$\beta$-$b$, as in the previous section.  

\begin{figure}
\centerline{\psfig{figure=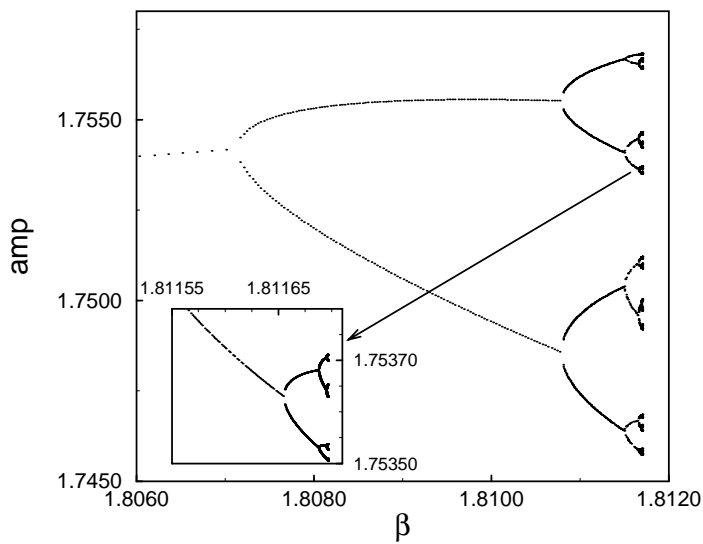,height=7.2cm  }}
\caption{A sequence of period doubling bifurcations for a network with
a $\sin$ transfer function and $N=3$. 
The weights consist of a single 
Fourier component without phase and bias. 
The vertical axis, denoted by `amp', is
the actual amplitude value of the cycle in phase space.
Insert:  a blowup of the pointed region.}
\label{fig_bif_N3}
\end{figure}

Before we turn to the analysis of the parameter space, let us demonstrate 
a manifestation of a chaotic behavior for concrete parameter values in
small networks, $N=3,4~$, via the mechanism of period doubling.
The dynamic system is described by Eq.\ (\ref{dyn_rule}) where the 
output value is given by Eq.\ (\ref{s_out}) with $~f=\sin$ transfer
function. Figure \ref{fig_bif_N3} presents a sequence of bifurcations of the 
output values (denoted by ``amplitude'') on a limit cycle 
as a function of the gain $\beta$ for $N=3$. 
For clarity we plot the sequence originating from one branch. 

% Figure 4 here

\begin{figure}
\centerline{\psfig{figure=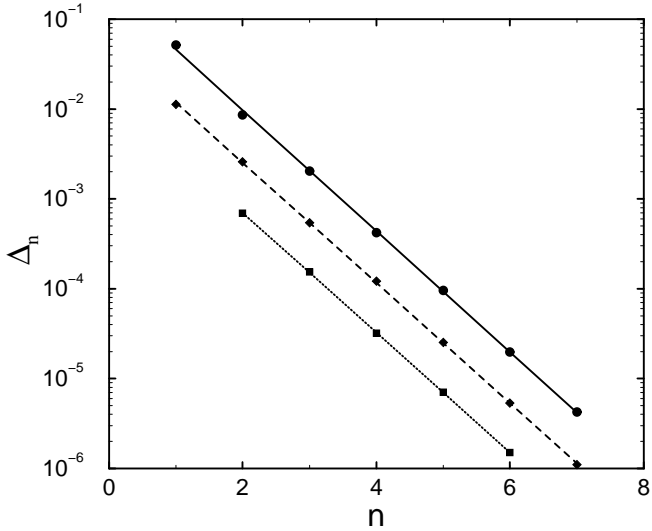,height=7.2cm  }}
\caption{The difference $\Delta_n$ (Eq.\ (\ref{delta_n})) 
for three cases. The
lines are exponential fit of the data. The solid line represents the fit
for $N=3$ with a slope of $4.69 \pm 0.05$. The dashed line represents $N=4$
with a slope of $4.66 \pm 0.02$ and the dotted line represents $N=4$ with
an arbitrary weights with a slope of $4.65 \pm 0.02$ .}
\label{fig_delta_bif}
\end{figure}

Figure \ref{fig_delta_bif} presents the 
difference between the values of $\beta$ at which successive period 
doubling occurs

\beq\label{delta_n}
\Delta_n= \beta_{n+1} - \beta_n
\eeq
\noindent
This figure depicts three cases: $N=3$, $N=4$ with weights consisting of a 
single Fourier component and $N=4$ with arbitrary weights.
Although the running index $n$ starts from $1$, the actual number of 
bifurcations is somewhat higher. Clearly the difference $\Delta_n$ 
is an exponential decreasing function of the form
\beq
\Delta_n \sim \delta^{-n}
\eeq
\noindent
The constant $\delta$ was evaluated from the slope for the three cases 
and found to be in good agreement with Feigenbaum's universal 
constant ($\sim 4.669$) \cite{feigenbaum}. \\

% Figure 5 here

The analytical analysis of the parameter space concentrates on obtaining 
the spine loci of a given map. The spine locus is associated
with the parameter vectors which lead to a super-stable attractor (e.g.\, in
one dimension, the first order derivatives of the map vanish at the 
super-stable attractor).
Barreto et.\ al.\ \cite{barreto} have conjectured that the structure of the
parameter space is determined primarily by the location and dimension of the 
spine loci. A window is constructed around the spine locus which leads to a
stable attractor. Generally speaking, the window is called `limited' if the 
spine locus is an isolated point in parameter space, whereas it is called 
`extended' if the spine is of higher dimension.

We turn now to a more systematic investigation of the parameter space,
starting with the simplest case of $N=2$. 
Since the number of free parameters is equal to the size of the system,
the number of positive exponents is at most the number of parameters, 
therefore one should not expect a robust chaos (unless fixing one of the
parameters).
The weights for a single Fourier component with $a=1$ and $\phi=0$
are given by Eq.\ (\ref{weights_1p}) . 
In principle we could take $\phi \neq 0$ which may drive a fraction of
the periodic orbits to quasi-periodic ones.

Similarly to Eq.\ (\ref{eq_sgen_N2}), the map can be written as follows
\beq\label{eq_sin_N2}
S^{t+1}= \sin \left[ \beta \left( (-1+b) S^t  + (1+b) S^{t-1} \right) \right]  
\eeq

\noindent
with the following fixed point (f.p.)
\beq\label{N2_fp}
S^{\star}=\sin(2 \beta b S^{\star}) 
\eeq

\noindent
The stability of the f.p.\  can be analyzed from its corresponding Jacobian
matrix
\beq\label{sin_jacobian_N2}
{\bf M}  = \left(	
\begin{array}{cc}
   \beta (-1+b) \cos(2 \beta b S^{\star})  &  
\beta (1+b) \cos(2 \beta b S^{\star})    \nonumber \\
1 & 0 
\end{array} 
\right)
\eeq

In order to identify the spine locus, the following
conditions must be satisfied: $~\det M = tr M = 0$ $~(\lambda_1=\lambda_2=0)$. 
It turns out that both constraints on the 
eigenvalues give rise to the same condition, $~\cos(2 \beta b S^{\star})=0$.
Therefore, we can say that the constraints are degenerate. Combining this
condition with Eq.\ (\ref{N2_fp}) we obtain $~S^{\star}~$ and the relation

\beq\label{N2_fp_curve}
\beta b = {\pi \over 4} (2 n + 1) \qquad n=0,1,\dots
\eeq

\noindent
This equation holds for $b>0$. For $b<0$ there are no f.p.\ solutions
that satisfy the constraint. 
The main spine for $b<0$ is related to a 2-cycle
solution which can be obtained numerically.	
Other f.p's which do not meet the constraint are possible and belong to 
a different curve in parameter space.

In a similar way, the constraints of the 2-cycle (~$S^{t+2}=S^t$~) 
spine locus gives the following relation between $\beta$ and $b$

\begin{eqnarray}\label{spine_locus_N2p2}
\sin (\beta \left[ (-1+b)+ (1+b)C \right] )  =  C \qquad \qquad \nonumber \\
C  =  {(4n+1)\pi-2 \beta (1+b) \over 2 \beta (-1+b)}  ~ , ~~ 
n=0, \pm 1, \ldots
\end{eqnarray}
\noindent

% Figure 6 here

\begin{figure}
\centerline{\psfig{figure=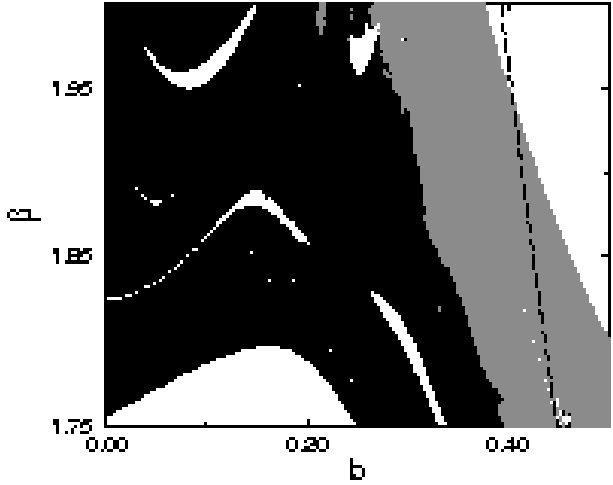,height=7.2cm  }}
\caption{Analysis of a region in parameter space for a network with
a $\sin$ transfer function and $N=2$ where points that lead to chaotic 
trajectories are marked. 
The dark(gray) colors correspond to areas with one(two) positive exponent.
The remaining space in this region leads to stable attractors. 
The bold dashed line is the
spine locus of the f.p.\ defined by Eq.\ (\ref{N2_fp_curve}), $n=0$.
}
\label{fig_sin_N2_general}
\end{figure}

\begin{figure}
\centerline{\psfig{figure=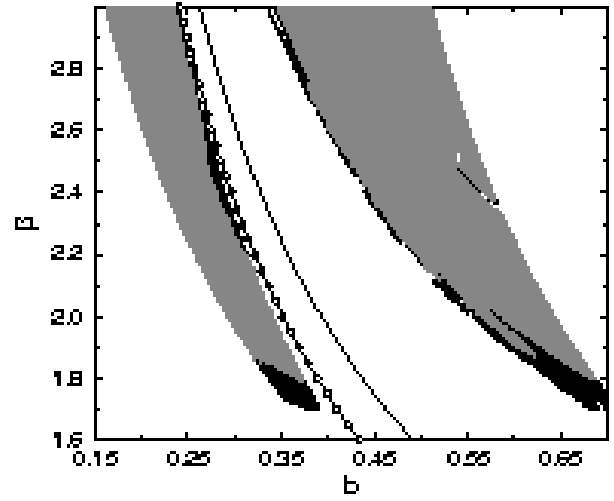,height=10.2cm  }}
\caption{Analysis of a region in parameter space around the main spine loci
for a $\sin$ transfer function and $N=2$. 
The initial condition in phase space is fixed 
to $S_i=1$. The dark(gray) color corresponds to areas with one(two) 
positive exponents.
The remaining space in this region leads to stable attractors. 
The solid line is the
spine locus of the f.p.\ defined by Eq.\ (\ref{N2_fp_curve}),
and the solid line with circles is the calculated
spine locus of the 2-cycle attractor defined by Eq.\ 
(\ref{spine_locus_N2p2}).
}
\label{fig_sin_N2_fpic}
\end{figure}

Turning to numerical analysis, Fig.\ \ref{fig_sin_N2_general} depicts 
a region of parameter space where areas
that lead to chaos are marked. This region was sampled exhaustively 
in a resolution of $~ \approx 10^{-5} ~$ in each direction. Several random 
initial conditions were used for each parameter value to avoid isolated 
cycles. The dark area corresponds to a region with
one positive exponent while the gray area corresponds to a region with
two positive exponents. The dashed line is the calculated spine locus of the
f.p.\ defined by Eq.\ (\ref{N2_fp_curve}) for the first branch, $n=0$.
Let us discuss briefly the structure of the parameter space.
The dark region (left-hand-side of the figure) contains 
extended (stable) windows (embedded white areas)
associated with cycles of different length.
The common feature of these windows is the fact that they are surrounded
by unstable regions with one positive exponent. 
As we move to the right-hand-side of the figure, a region with two positive 
exponents emerges. 
The spine locus depicted enters this region since, as mentioned above, we
started from several initial conditions, therefore the dynamics is typically
attracted to unstable cycles of higher order. 
In order to isolate the spine of the f.p.\ and the 2-cycle from higher order
cycles, we analyze the window 
when the initial condition is fixed to $S_i=1$.
Figure \ref{fig_sin_N2_fpic} depicts a region of parameter space for which
this analysis was applied.
In this case, the basic nature of these spines is revealed and a clear 
extended window is constructed around the solid curve 
(f.p., Eq.\ (\ref{N2_fp_curve})) as well 
as the solid curve with circles (2-cycle, Eq.\ (\ref{spine_locus_N2p2})). \\

Analysis of the case $N=3$ is similar to $N=2$. 
The spine locus for the f.p.\ is given by
\beq\label{fp_spine_N3}
\beta b = {(2n+1) \pi \over 6}  \qquad b>0 \quad , ~ n=0,1,\dots
\eeq
\noindent
We can generalize the equation for the spine locus of the f.p.\ for any $N$

\beq\label{fp_spine_general}
\beta b N = {(2n+1) \pi \over 2}  \qquad b>0 \quad , ~ n=0,1,\dots
\eeq

The constraint for the 2-cycle is obtained from similar conditions 
formulated for $N=2$.
The relation between $\beta$ and $b$ is the following

\begin{eqnarray}\label{spine_locus_N3p2}
\sin \left( \beta \left[ (1/2+2b)+ (-1/2+b)C \right] \right) =  C 
\qquad \nonumber \\
C =  {(4n+1)\pi + \beta (1-2b) \over \beta (1+4b)} ~ , ~~ 
n=0, \pm 1, \ldots
\end{eqnarray}

The case $N=3$ reveals another aspect in the structure of parameter space.
There are regions for which we find three positive Lyapunov exponents.
In such regions, we observed a robust chaos, namely, small changes of 
the parameters would not destroy the chaotic behavior. 

In principle we can construct the conditions of the spine locus for larger 
cycles and larger systems, however the task becomes much more involved 
as the cycle length increases. \\

Let us now extend our analysis for large systems. Two questions come to the
fore: 
\begin{enumerate}
\item Can we find regions for which the chaotic dynamics is robust, and how
frequent are they ? 
\item Is there a simple relation between the attractor 
dimension and the control parameters or, in other words, can we control the  
attractor dimension ?
\end{enumerate}

We saw previously that even in the case $N=3$, a robust chaos is
observed. However, this type of dynamics is of little interest since the
volume in phase space is expanding, $~\sum_{i=1}^N \lambda_i > 0~$, hence
the bounded space is filled. The more interesting case is a motion which
is confined to an attractor, yet the number of positive exponents is larger
than the number of free parameters. We claim that the possibility to find
regions with an increasing number of positive exponents, grows with $\beta$.
This means that we have a natural parameter in the model that controls the 
degree of the chaos. In addition, this parameter controls the dimension of the
attractor.

In order to test this hypothesis we used a larger system, $N=17$. 
To convince the reader that our analysis is not
restricted to the simple case of a single Fourier component, we used 
more complicated weights consisting of two Fourier components 
with irrational phases and a bias term. 
The amplitudes and phases of the components were kept fixed, 
therefore we have the same two dimensional parameter space, as before. 
The exact details of the amplitudes and phases are of no importance. 

A close inspection of the parameter space reveals the following regimes:
First, the incommensurate regime which corresponds to the irrational 
phases of the weights. 
Above some value of the gain parameter (depending on the details of
the weights) most of the space is associated with chaotic dynamics. The number
of positive exponents in this regime grows as $\beta$ increases until the 
sum of the exponents becomes positive. In this regime, we observed a 
relatively monotonic growth in the attractor dimension, calculated using 
Eq.\ (\ref{kap_yorke}) (see Appendix). Figure \ref{fig_ad_N17_sin} 
depicts the attractor 
dimension for a fixed bias value. Clearly, the dimension grows 
monotonically. The figure also shows the number of positive exponents 
which grows with $\beta$. 
Each point was averaged over $10$
random initial conditions in order to check whether the same attractor is 
sampled. Indeed, the errors are less than $1 \%$ and typically much less,
therefore they are not presented.
(Note that there are cases, not shown, for which the line $b=\mbox{const}$ 
crosses a window.
In such regions, the attractor dimension decreases and then continues to grow
once the window is passed). As the sum of the exponents becomes positive, the 
attractor dimension saturates the dimension of the system, $N$. 

% figure 7 here

Finally we validated this results using a different method. 
We tested several points in parameter space 
by estimating the attractor dimension from 
the time series generated by a network and compare it to the estimation
using Eq.\ (\ref{kap_yorke}).
The time series was recorded from a system with 
the same parameters and the attractor dimension was calculated from the
reconstructed phase space using the method of Correlation-Integral 
\cite{grassberger}. The results confirm our hypothesis for the monotonic 
relation between the number of positive exponents and $\beta$.

\section{Combination of monotonic and non-monotonic functions}\label{sec_comb}

In this section we discuss the mixed case where the transfer
function can be written in the following way

\beq\label{fun_comb}
f(x)=f_{m}(x) + \epsilon f_{nm}(x)
\eeq

\noindent
where $f_{m~(nm)}$ represents a monotonic (non-monotonic) function; 
$\epsilon$ is a mixing parameter (not necessarily small). For concreteness,
assume that $~f_{m}(x)=\tanh (x)~$ and $~ f_{nm}(x)=\sin (x)$. 
Let $~x= \beta {\vec W}\cdot{\vec S}~$ and the weights are given by Eq.\
(\ref{weights_1p}) (taking $\phi=0$ for simplicity). Following the same 
developments shown in \cite{prl,nips} we develop an asymptotic 
periodic solution of the form

\beq\label{comb_sol}
S_l = \tanh ( A \cos({2\pi \over N} k l + B ))+ 
      \epsilon \sin ( A \cos({2\pi \over N} k l + B )) 
\eeq

\noindent
where  $~A=A(\beta) ~,~B=B(\beta)$. The coefficients $A,B$ can be obtained 
from the self-consistent equations :

\begin{equation}\label{sol_1p_var}
\begin{array}{cclcc}
A & = &  {1 \over 2} \beta N a \sum_{\rho=1}^{\infty} D(\rho) (A/2)^{2\rho-1} 
  (\rho !)^{-2} & ; &  B=0  \nonumber \\ \nonumber \\
B & = & \beta N b \sum_{\rho=1}^{\infty} D(\rho) 
   B^{2\rho-1} ((2\rho)!)^{-1}  & ; & A=0 
\end{array}
\end{equation}

\noindent      
where $~D(\rho)= {2^{2\rho}(2^{2\rho}-1) } {\cal{B}}_{2\rho} + 
2 \epsilon \rho (-1)^{\rho+1} ~$
and ${\cal{B}}_{2\rho}~$ are the Bernoulli numbers.

As described in \cite{nips}, when the gain value increases
the system undergoes a transition from the 
trivial solution, $~S_l=0$, to a state which is governed by one of the two
possible attractors: a fixed point ($A=0~,~B \neq 0$) or a periodic solution,
depending on the relation between $~\beta_{c1},\beta_{c2}$ (the critical value
for the onset of each attractor). At higher gain values 
($\beta> \beta_{c1},\beta_{c2}$) both attractors are stable and the system will
flow to one of them, depending on the initial condition. 
When the gain parameter is further increased, one observes unstable dynamics of
the type described in this paper. The mixing parameter $\epsilon$ controls
the actual point from which the parameter space is governed by the 
non-monotonic function.

Finally, we note that it is easy to generalize this solution for $\phi \neq 0$
and weights which contain more Fourier components 
(Eq.\ (\ref{weights_general}) following \cite{prl,nips}).

\section{Discussion}\label{discussion}

In this paper we analyzed a class of neural networks in the context of
time series generation and focused on the effect of the transfer function
on long-term behavior. 
We suggested a natural way to classify transfer functions 
into monotonic and non-monotonic functions. The class of monotonic 
functions can generate chaotic dynamics, however the weights should
contain a non-trivial bias term and a phase. The chaos
can be regarded as fragile, i.e.\, in regions where unstable behavior can be 
observed, the parameter space is characterized by points 
around which the parameter vectors lead to a stable dynamics 
(limited windows). As the size of the network increases, the structure of
parameter space becomes more involved due to the appearance of longer cycles.
On the other hand, the class of non-monotonic functions
is capable of generating robust high dimensional (chaotic) attractors.
This means that there
exist regions of parameter space for which slight changes in the vector 
of accessible parameters will not stabilize the system.
Although the analysis presented in the paper was exemplified by an odd 
non-monotonic function, the results we obtain, which are mainly related to the 
regime of $\beta$ values where unstable behavior emerges, include
even functions as well \cite{long_ver}.

One must refer to another paper \cite{ICNN96} that focuses on
searching robust chaos in recurrent neural networks using weight space
exploration. Their motivation and methods differed from ours. However,
their main result which states that robust chaos can be achieved using a 
non-monotonic transfer function only, is in agreement with our conclusions.\\

Another aspect characterizing non-monotonic functions is the potential 
to monotonically increase the attractor dimension over a broad range 
of parameters. This interesting effect is achieved by increasing the gain 
parameter, $\beta$. Unless the path of the vector of parameters
intersects the boundary of a spine locus, the attractor dimension increases
monotonically with $\beta$ until it saturates the dimension of the system.
From this point, one can no longer define the dynamics as an attractor 
since a volume of phase space expands.
 
Based on these results we can formulate the following conclusion: 
A `perceptron-SGen' with non-monotonic
transfer function can generate a chaotic attractor much more easily
than with a monotonic function.
In addition, when the chaos is robust one can expect the learning process
to be easier since the extended parameter space which includes the attractor
dimension is smooth in these regions.

We further showed that the results can be extended to transfer functions
which are combinations of monotonic and non-monotonic functions. 
The asymptotic stable attractor is developed similarly to the pure case of
monotonic / non-monotonic functions. The structure of the unstable region 
is governed by the function that looses its stability in that region. There
are three possible scenarios: either one of the two types of functions loses
its stability while the other remains stable or both functions lose their
stability. The first two cases are actually covered throughout the paper. \\

The extension of the results to the case of Multi-Layer Networks raises
new interesting questions, e.g.\,
to what extent our findings remain valid; how does
the combined perceptron-SGen's affect each other, namely, do they act to
stabilize the dynamics or does a chaotic unit maintain its behavior, thus 
reflecting its robust nature.
Another issue of major importance is the 
attractor dimension of the MLN. While a `perceptron-SGen' can typically 
generate a $1D$ attractor in the stable regime, we saw that applying a 
non-monotonic transfer function to this network can generate much higher and
continuous attractor dimension. On the other hand, a `MLN-SGen' can generate
higher (integer) attractor dimension in the stable regime 
(see \cite{prl,nips,sgen_mln}). Do `MLN-SGen's generate continuous high
dimensional attractors as well ? 

\section*{Acknowledgments}
We thank W. Kinzel and Y. Ashkenazy for fruitful discussions. 
I. K. acknowledge the support of the Israel Academy of Sciences.

\section*{Appendix}\label{appendix}

In order to characterize the parameter space numerically, we apply a rather 
straightforward method. For a dense mesh of points in the $\beta$-$b$ plane
we calculate the spectrum of Lyapunov exponents associated with the dynamics,
from which we can obtain the stability of the attractor and its attractor
dimension. This procedure is applied after the transient and for several
randomly chosen initial conditions.
The spectrum is estimated using an algorithm suggested by Wolf
et.\ al.\ \cite{wolf_alg}. 
Basically, the algorithm evolves an orthonormal basis by multiplying each
vector with the Jacobian matrix  which
is evaluated along the trajectory. To overcome the problem
of exponential decreasing of the vectors associated with smaller eigenvalues, 
the principal vectors are re-orthonormalized frequently. This procedure 
ensures that our analysis does not run into roundoff errors.
This algorithm measures the average exponential
change of a volume along the trajectory in state-space, using the 
rate of change in the principal vectors. 

The attractor dimension is estimated
using the Kaplan-Yorke conjecture \cite{kap_yorke} that gives the following
relation between the (sorted) spectrum of Lyapunov exponents $\lambda_i$ 
($\lambda_1 > \lambda_2 \ldots $)
and the attractor dimension $d_{ky}$ (information dimension)
\beq\label{kap_yorke}
d_{ky} = n + \frac{\sum_{i=1}^{n} \lambda_i}{| \lambda_{n+1} |}
\eeq
\noindent
where $n$ is defined by the condition $\sum_{i=1}^{n} \lambda_i > 0$ and
$\sum_{i=1}^{n+1} \lambda_i < 0$. 

\noindent
We are aware of a problem associated with the Kaplan-Yorke conjecture. 
In cases where the spectrum has some very small exponents it is possible 
to obtain a biased estimation of the dimension since the sum described 
in Eq.\ (\ref{kap_yorke}) may fluctuate around zero due to one of the 
exponents. In such cases, we take additional precaution to avoid this problem.

%\newpage

%\section*{References}

\newpage

% **** FIGURES *****

%\vspace{3.cm}

\begin{figure}
\centerline{\psfig{figure=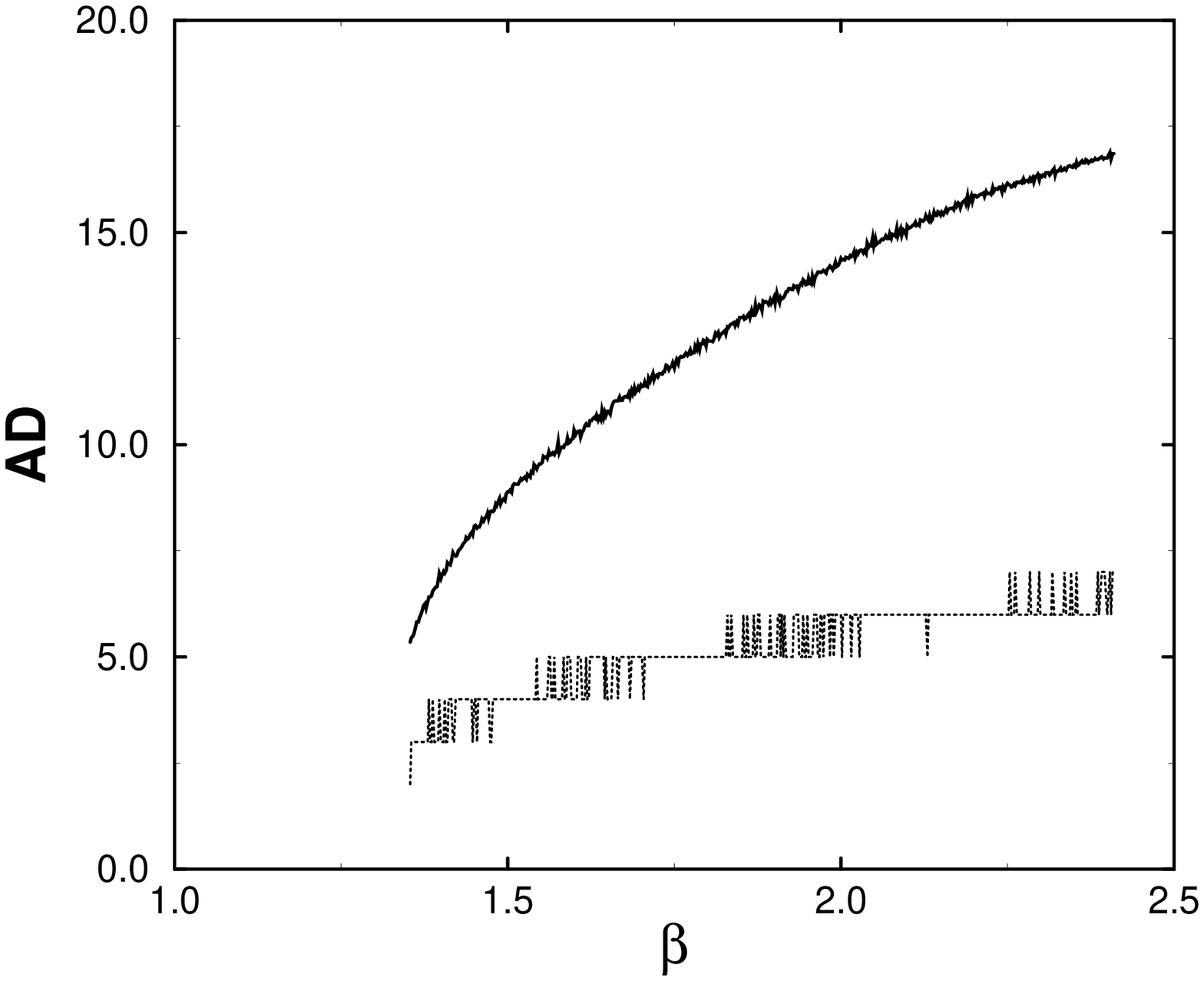,height=7.2cm  }}
\caption{The attractor dimension (AD) as a function of the gain
for a $\sin$ transfer function where $N=17$, $b=0$ and $\phi=0$.
The solid curve is the AD and the dashed line below 
represents the number of positive exponents. 
}
\label{fig_ad_N17_sin}
\end{figure}


\begin{thebibliography}{99}
\bibitem{annni_explain}
An Ising system on a Cayley tree with infinite coordination limit and
competing interactions between first (FM) and second (AFM) neighbors 
along an axial direction, is represented by a second order recursion 
relation for the magnetization per spin as Eq.\ (\ref{eq_sgen_N2}).
\end{thebibliography}
\end{document}